\documentclass{eas}
\usepackage{graphicx}
\begin{document}
\title{The Formation \& Evolution of Galaxies making up the CIRB : FIR/Submm Extragalactic Surveys from Dome C}
\author{Mattia Vaccari}
\address{Department of Astronomy, University of Padova, Vicolo Osservatorio 3, I-35122, Padova, Italia}
\secondaddress{\texttt{mattia@mattiavaccari.net}}
\author{Alberto Franceschini}\sameaddress{1}
\begin{abstract}
The science case for FIR/Submm surveys of the extragalactic sky to be carried from Dome C are reviewed. The main questions concerning the formation and evolution of galaxies making up the CIRB are outlined and opportunities to exploit Dome C unique observing conditions through single-dish observations are discussed.
\end{abstract}
\maketitle
\section{FIR/Submm Extragalactic Astronomy}
The wide spectral region between the optical and radio windows has enjoyed an increasing popularity ever since the IRAS mission (\cite{Neugebauer1984}) opened up the 10-100 $\mu$m wavelength range to a systematic investigation in 1983. In its short lifetime the satellite discovered a sizeable population of galaxies emitting more than 95\,\% of their total luminosity in the infrared, showed that the starburst phenomenon is ubiquitous, and that starburst galaxies must undergo an extreme evolution with redshift if they are to accounf for the observed Cosmic Infrared Background (CIRB). The first homogeneous all-sky survey ever, IRAS has not only had a strong impact on both Galactic (\cite{Beichman1987}) and extragalactic (\cite{Soifer1987}), but also demonstrated how image atlases and source catalogs of outstanding quality (\cite{Beichman1988}) can have a long-lasting impact.

Over the following 25 years, our knowledge of the properties of infrared galaxies and their relationship with optical galaxies has improved by leaps and bounds, helped by wide-ranging technological breakthroughs as much as by high-altitude ground-based observatories, air-borne, baloon-borne, or satellite-borne instrumentatin, gradually overcoming the hardships imposed by atmospheric absorption. In this context, the loosely defined FIR/Submm wavelength range (30-1000 $\mu$m) has played a central role in two of the most striking intervened discoveries, namely the discovery of an isotropic 100-500 $\mu$m background of likely cosmological origins with COBE in 1996 and SCUBA 850 $\mu$m blank-field surveys implying sustanined star formation activity well above a redshift of 2 in 1998. 

\begin{figure}
\centering
\includegraphics*[width=\textwidth]{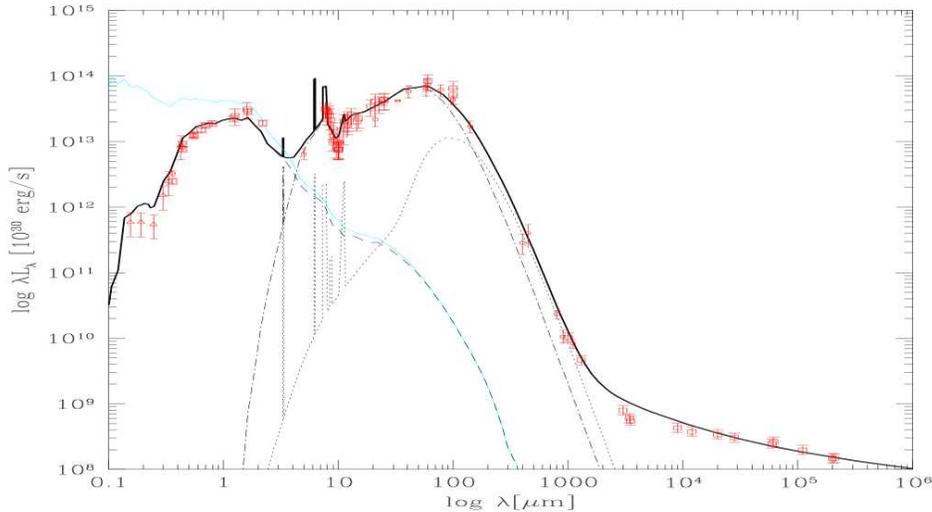}
\caption{Broad-band (UV through radio) SED of prototype local starburst  galaxy M82.}
\label{m82-bb-spectrum.fig}
\end{figure}

\begin{figure}
\centering
\includegraphics*[width=\textwidth]{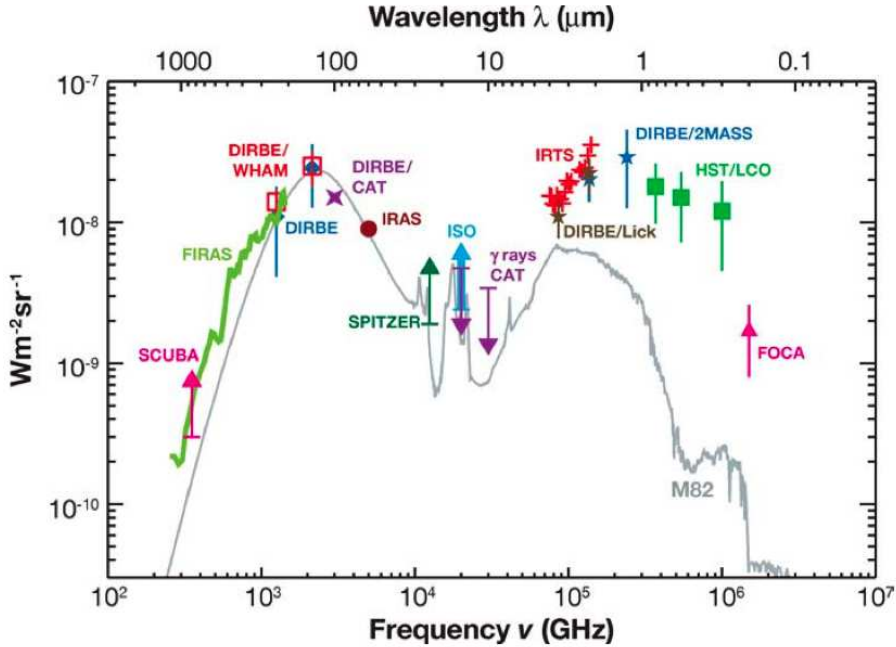}
\caption{Broad-band cosmic background spectrum, superposed on M82 SED. The FIR/Submm peak, enclosing close to 50\,\% of the overall energy, lies at about 200 $\mu$m. From \cite{Lagache2005}.}
\label{cirb-spectrum.fig}
\end{figure}

These two elements, confirmed by several other pieces of evidence, strengthened the importance of the FIR/Submm in disentangling the star formation history of the Universe up to the highest redshifts and in evaluating the relative importance of black hole accretion and star formation on the overall cosmic energy budget. This is because the SEDs of bolometrically-bright high-redshift starbursts are well-described by local starburst spectral templates such as M82, whose emission peaks at 30-100 $\mu$m
(See Figure~\ref{m82-bb-spectrum.fig}). This, combined with an abundant high-redshift starburst population, causes the CIRB to peak at about 200 $\mu$m (See Figure~\ref{cirb-spectrum.fig}), with most of the highest-redshift sources still observable at longer wavelengths  due to the favourable $K$-correction setting in.
\begin{figure}
\centering
\includegraphics*[width=\textwidth]{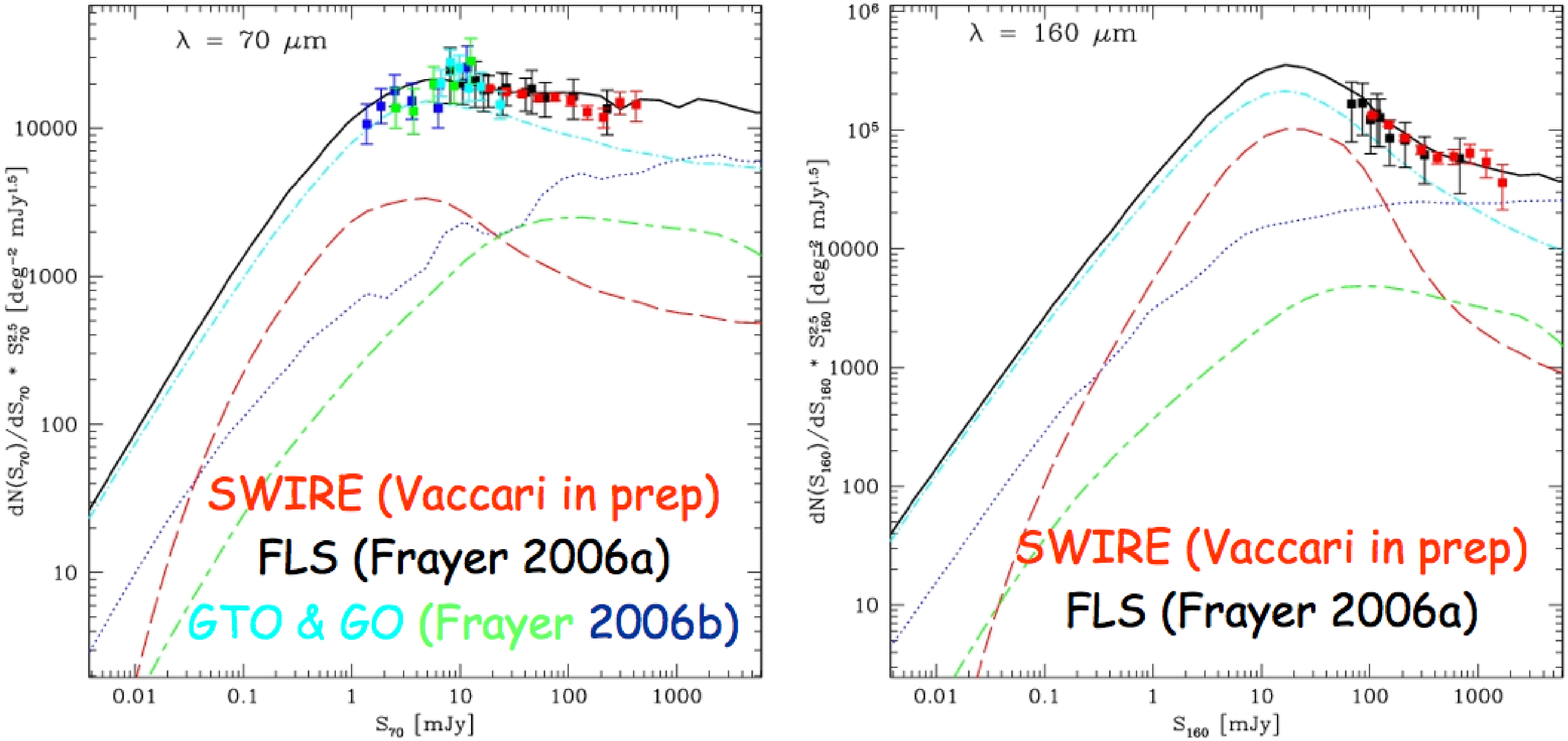}
\caption{70 \& 160 $\mu$m differential counts from several Spitzer observing programs, superposed on models for long-wavelength extragalactic populations by \cite{Franceschini2008}. Coloured lines are the individual populations while the black line is the total. The upturn in the counts is interpreted as a strong evolution with redshift in the number and the luminosity of the populations. From \cite{Vaccari2008}}
\label{mips-ge-counts.fig}
\end{figure}

In the process, a few questions have emerged and largely remained unanswered, e.g. the origin of massive ellipticals observed at least up to $z\sim2$, the mechanisms regulating the co-evolution of black holes and their host galaxies and the role of merging. Over the last decade, through coordinated observing programs of the most popular cosmological fields, the ISO, Spitzer \& Akari satellites have contributed to shed new light on these issues, although the limited angular resolution provided by their $< 1$\,m telescope largely confined them to wavelengths shorter than 100 $\mu$m. It will thus be left to Herschel, the first 4-m class telescope in space whose launch is expected in early 2009, to open up the 70-500 $\mu$m range to photometric surveys over large areas and, to a more limited extent, to spectroscopic investigation at roughly the same wavelengths. The SPICA mission, if approved, will then leverage the advantage of an Herschel-size cold telescope by carrying out deep spectroscopy over the 30-200 $mu$m range, while JWST will do so over the complementary 5-30 $\mu$m range. A number of ground-based Submm dishes will complement these efforts with longer-wavelength observations, building on pioneering efforts by CSO, SCUBA \& MAMBO.  ALMA will finally provide outstanding interferometric capabilities throughout the 0.3-10\,mm range, althoughits mapping capabilities will be somewhat limited and will therefore mainly be used as a spectroscopic follow-up instrument. One must then ask, guided by state-of-the-art models reproducing the statistics provided by previous observing efforts (See e.g. Figure~\ref{mips-ge-counts.fig}), whether a Dome C FIR/Submm observatory may be a competitive player in this crowded field. In so doing, we refer to the proposed design for an ALMA-like winterized 12\,m antenna (\cite{Olmi2007,Olmi2008}) equipped with a 10,000-pixel filled bolometer array operating at 200/350/450 $\mu$m and characterized by the parameters reported in Table~\ref{sens.tab}. These assumptions represent realistic estimates of NEFD ranges to be expected at Dome C and will be used by the ARENA working group established to carry out a detailed assessment of FIR/Submm science cases throughout 2008. This instrumental configuration will be hereafter referred to as the Antarctic Submm Observatory (ASO) concept.
\begin{table}
\centering
\caption{Instrumental parameters for an ALMA-like winterized 12\,m antenna equipped with a 10,000-pixel filled bolometer array operating at 200/300/450 $\mu$m, or the ASO concept. The NEFD ranges roughly correspond to the expected 25\,\% and 75\,\% PWV percentiles and on realistic assumptions for telescope aperture efficiency and optical transmission. and bolometer absorption. Courtesy of Vincet Minier.}
\label{sens.tab}
\smallskip
\begin{tabular}{|c|c|c|c|}
\hline
$\lambda$ [$\mu$m & NEFD [mJy/beam] & Beam ($\lambda/D$) [arcsec] & FOV [arcmin$^2$] \\
\hline
200 $\mu$m &  500--1000 & 3.4  & $2.8\times2.8$ \\
350 $\mu$m &  100--200 & 6.0  & $5.0\times5.0$ \\
450 $\mu$m &  100--200 & 7.8  & $6.5\times6.5$ \\
\hline
\end{tabular}
\end{table}
\section{Atmospheric Transparency at Dome C}
Interest in Dome C as a FIR/Submm observing site owes much to its atmospheric transparency at these wavelengths. While we still lack a consistent collation of FIR/Submm site testing data, Dome C seems to provide a clear advantage throughout the 200, 350 and 450 $\mu$m atmospheric windows, although concerns about both the stability and the low overall levels of atmospheric transmission still stand (\cite{Minier2007,Minier2008}). The 200 $\mu$m window appears particularly interesting over the long term, since this is where Herschel \& SPICA sensitivity will be hampered by confusion due to their $<4$\,m telescope size, and where the advantage over Chilean Submm sites seems more compelling. 
\section{Extragalactic Confusion Limits}
When instrumental and natural backgrounds are kept down to manageable levels, the sensitivity of FIR/Submm observations soon becomes limited by the sky background caused both by structure in the solar system and in the Galaxy and by an abundant population of point-like extragalactic sources. While several components to this "noise" must in general be considered, for deep cosmological surveys of high-latitude fields the contribution from extragalactic sources, also knows as extragalactic confusion, is by far the dominant one.

The flux levels at which confusion sets in can be estimated on the basis of models for the properties of long-wavelength extragalactic populations such as those by \cite{Franceschini2008} used in Figure~\ref{mips-ge-counts.fig} using two complementary approaches. The first approach (hereafter labelled ÓFluctuationsÓ and discussed by \cite{Franceschini1989} follows from an analysis of the cell-to-cell fluctuations due to randomly distributed unresolved sources. This is carried out by modelling the telescope deflection probability distribution using a Gaussian beam profile with a given FWHM and then computing the rms signal ($\sigma$) due to sources below the confusion limit. The implicit relation between $\sigma$ and the FWHM is solved iteratively by fixing ? and deriving the FWHM. In our case, ÓconfusionÓ is assumed to set in at the 4\,$\sigma$ level. The second approach (hereafter labelled ÓCountsÓ, and discussed by \cite{Franceschini2001} follows from the maximum number of resolved sources to be found within each beam, and sets ÓconfusionÓ at the 30 beams / source level. Roughly speaking, while the former approach follows the trend of source counts fainter than the confusion limit, the latter follows from source counts brighter than that. The degree of consistency between the two approaches thus depends on the slope of source counts below the confusion limit, while providing a measure of the robustness of the derived estimates.

\begin{figure}
\centering
\includegraphics*[width=0.475\textwidth]{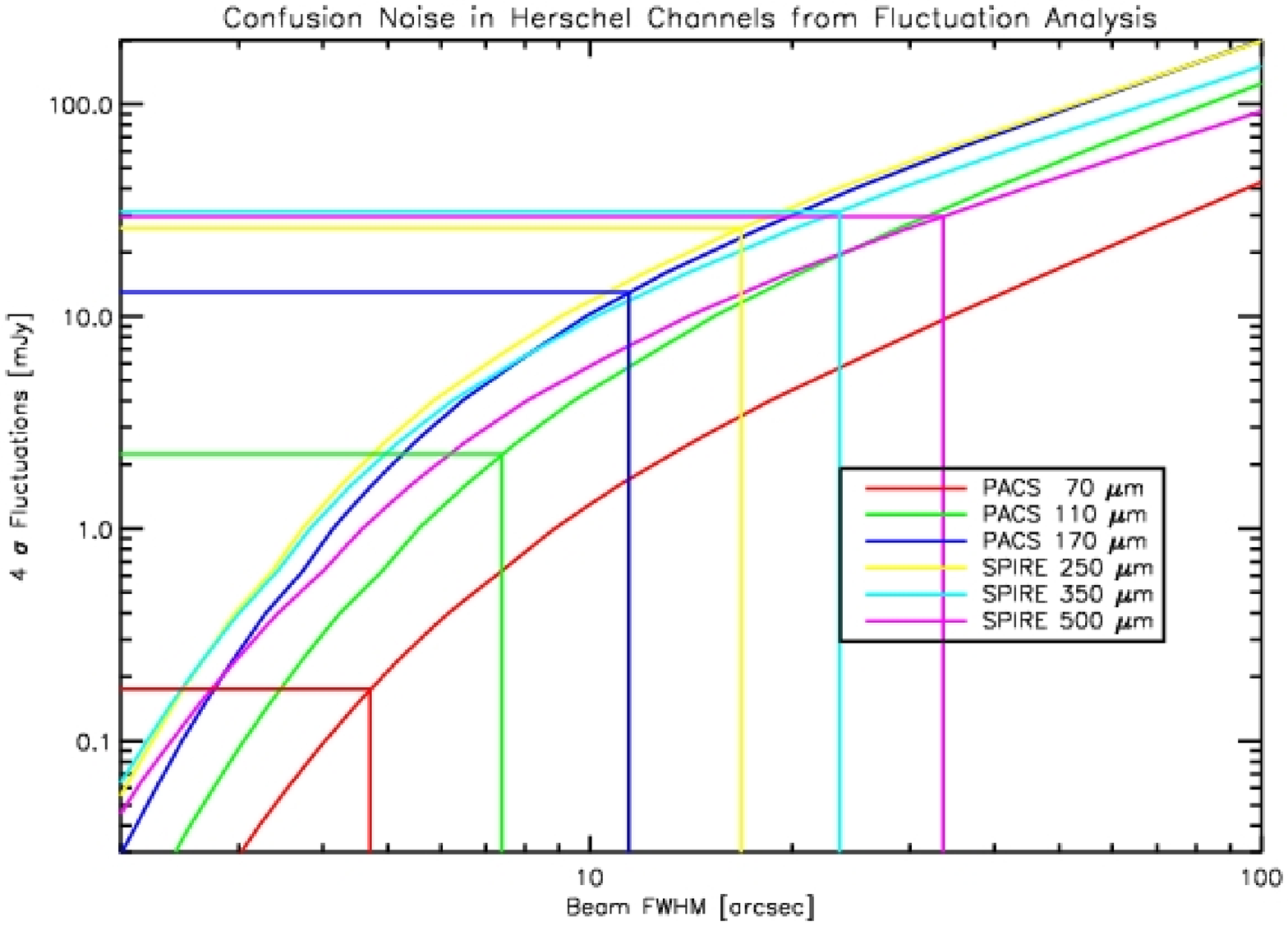}
\includegraphics*[width=0.475\textwidth]{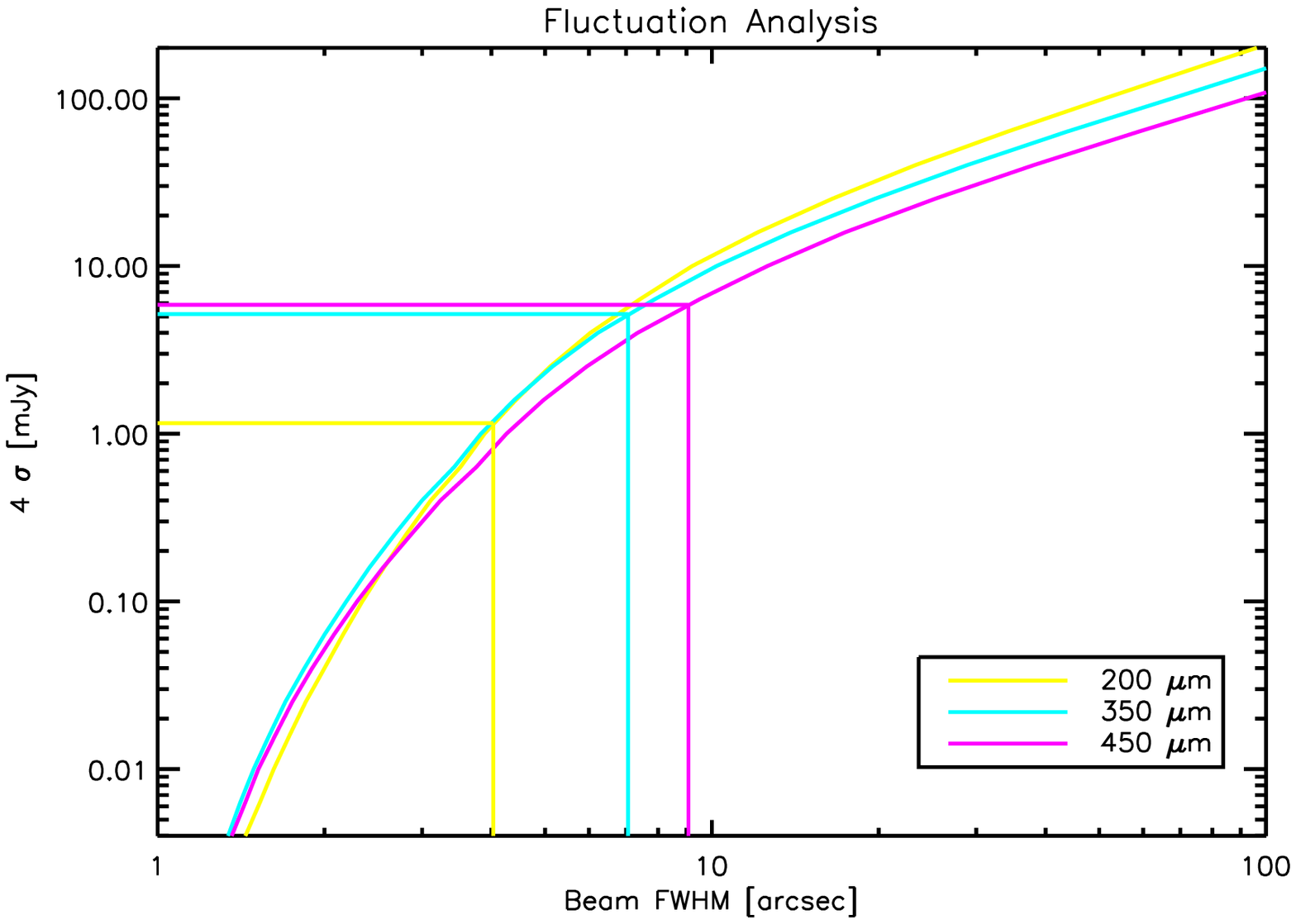}
\caption{Expected confusion limits for Herschel (left) and ASO (right) channels on the basis of the models by \cite{Franceschini2008}. Similar colours are used for similar wavelengths. See text for details.}
\label{confusion.fig}
\end{figure}

In the present FIR/Submm case, given the rather steep slope of the counts at the wavelengths and flux levels under consideration, we adopt the "Fluctuations" approach, which provides a more accurate (and conservative) estimate of confusion under the circumstances. Figure~\ref{confusion.fig} provides a comparison of expected confusion limits for Herschel (3.6\,m) and ASO (12\,m). The intersection of coloured curves describing fluctuation levels with coloured lines corresponding to beam sizes at various wavelengths provides an estimate of confusion limits. The limits for the 250/350/500 $mu$m Herschel bands consistently fall at fluxes of 30 mJy while ASO would be able to reach down to 5 mJy at 350 \& 450 $\mu$m and down to 1 mJy at 200 $\mu$m.

\begin{figure}
\centering
\includegraphics*[width=\textwidth]{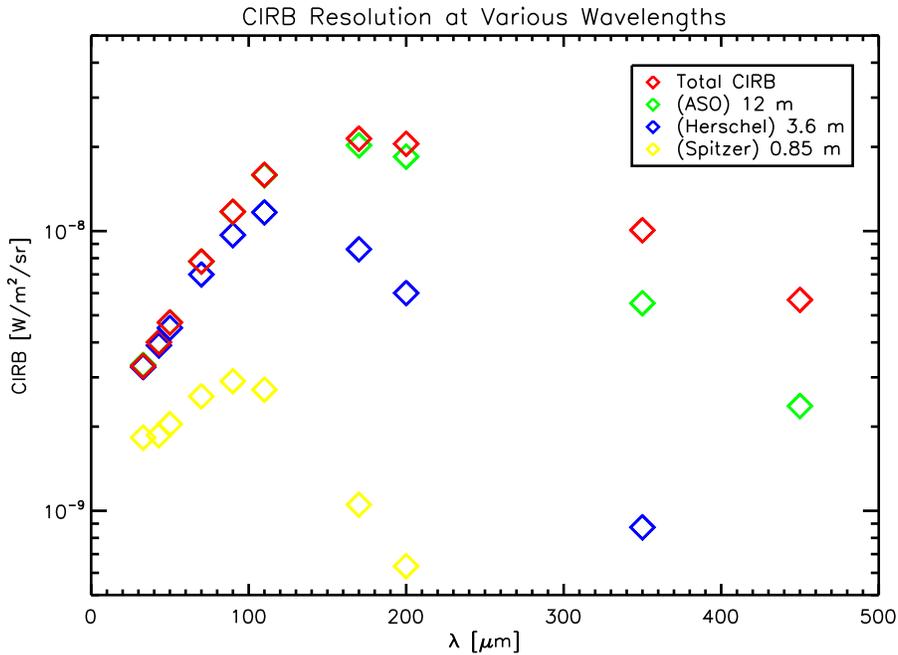}
\caption{CIRB Resolution from Ground \& Space. Percentage of CIRB resolved by Spitzer, Herschel \& ASO as a function of wavelength. See text for details.}
\label{cirb-all.fig}
\end{figure}

The above confusion estimates set fundamental limits on the detectability of fainter sources (although it is estimated that multi-wavelength source extraction techniques might push down this limit by a factor of $\sim 2$). It is therefore interesting to estimate which percentage of the integrated CIRB would be resolved into sources at each wavelength should we carry out confusion-limited surveys with ASO and other facilities. Figure~\ref{cirb-all.fig} details such a comparison, showing how ASO would be able to resolve virtually all of the 200 $\mu$m CIRB and about half of the 350 \& 450 $\mu$m CIRB into discrete sources, while the fraction resolved by Herschel (but also by SPICA, which if approved would share Herschel 3.6\,m mirror size) rapidly decreases with wavelength as we move above 100 $\mu$m.
\section{Practical Sensitivities}
As seen previously, Dome C therefore offers an attractive opportunity for deep observations at Submm wavelengths in atmospheric windows which very rarely open up at even the best astronomical observing sites. In order to fully exploit this opportunity for cosmological surveys, however, the sensitivity of the facility to be put in place at Dome C must be such that it allows to actually achieve the above confusion limits over practical observing times. Unfortunately, when compared to space instrumentation such as Herschel, ASO will suffer from a much higher natural background emission which will roughly compensate the advantage offered by its larger aperture than Herschel's. Using parameters from Table~\ref{sens.tab}, and namely the average NEFD values, the time required to cover 50\,arcmin$^2$ (or one third of each of the GOODS fields) down to 1 mJy/beam rms would be of order 1000 hr at 200 $\mu$m but only of order 10 hr for 350 \& 450 $\mu$m. Given that the 200 $\mu$m window will only open some of the time, the viability of the ASO concept for fully-fledged cosmological surveys at 200 $\mu$m appears to be questionable at the very least. The mapping speed, i.e. the time required to cover a given area down to a given flux, scales as $D^2$. Even more dramatically, the time required to reach a given flux limit scales as $D^4$, although the sharper PSF provided by a larger dish means than an array detector with a given number of elements will Nyquist-sample a smaller field of view, hence the more limited quadratic effect on the mapping speed. For this reason, while 350 and 450 $\mu$m surveys could still be routinely and effectively carried out, the cosmological impact of a 200 $\mu$m channel operating at a 12\,m antenna will be somewhat limited by its reduced sensitivity, and a substantially larger dish, e.g. the 25\,m currently envisaged by the CCAT project, would be required to suitably fulfill the promise of the 200 $\mu$m Dome C window. 
\section{Simulated FIR/Submm Surveys from Dome C}
\begin{figure}
\centering
\includegraphics*[width=\textwidth]{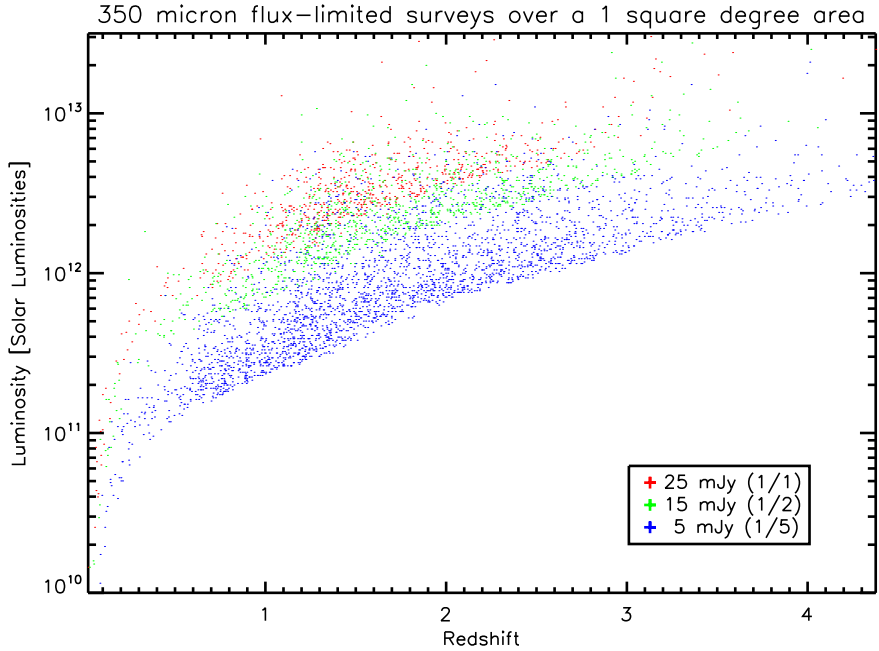}
\caption{Simulated 350 $\mu$m flux-limited surveys over a 1 deg$^2$ area. Three different flux ranges are chosen, roughly representing Herschel confusion limit (25 mJy), Herschel ultra-deep survey limit and ASO confusion limit (5 mJy). For clarity, only one in two sources is plotted at 15 mJy and one in five at 5 mJy.}
\label{350-micron-surveys.fig}
\end{figure}

Give the limitations to its sensitivity outlined above, it is probably helpful to currently think of the ASO concept as a 350 \& 450 $\mu$m surveyor with 200 $\mu$m follow-up capabilities. According to Table~\ref{sens.tab}, mapping 1\,deg$^2$ down to 1 mJy (i.e. a $5 \sigma$ limit equal to the confusion limit of 5 mJy) would take ASO about 900 \& 500 hr at 350 \& 450 micron, respectively. At a speculative 18 observing hr per day, such a massive 1,400 hr program could be accomplished within a mere three months if, as suggested by Figure~3 of \cite{Minier2007}, Dome C atmospheric transmission at 450 $\mu$m remains above 50\,\% for 90\,\% of the time. By itself, this programme would allow to image all ultra-deep fields targeted by Herschel Key Programmes with a much-improved angular resolution and down to a flux fainter by a factor of 2.5 or more. Figure~\ref{350-micron-surveys.fig} shows how this would allow not only to greatly reduce confusion and therefore multi-wavelength identification problems, but also reach down to below $10^12$ solar luminosities and up to redshifts above 2.5 to greatly expand the parameter space available for exploration.
%Similar gains would be obtained over smaller-area deeper surveys of larger-area shallower ones.
The 450 $\mu$m channel would then be even more efficient at detecting high-redshift galaxies.
\section{Practical Issues}
Independently of the size of the antenna to be deployed and of related engineering issues, a number of practical issues must be dealt with in order to optimize the science return of extragalactic surveys to be carried out by a Dome C FIR/Submm antenna, and in particular:
\begin{itemize}
\item \textbf{The Observatory vs. Facility Instrument Issue} : the difficulties and the peculiarities inherent in operating a large telescope in Antarctica call for an effort of simplification of its operations. On the other hand, gathering support for such an endeavour will have more chances to succeed if the project is not perceived as closed but rather as open to a proposal process. Similarly, while pointed observations lined up in some sort of observing queue might be desirable for the community at large, a survey instrument with a few rock-solid observing modes would guarantee smoother operations and thus less pressure on winter-over astronomers.  Bearing this in mind, community involvement in and operational scenarios of the observatory will have to be carefully devised and discussed.
hshsjjsjs
\item \textbf{The Field Choice Issue} : historically, most extragalactic surveys have been carried out at moderate latitudes, and there lie most "Cosmic Windows", i.e. sky areas targeted over the last decade or so by cosmological surveys undertaken at "all" wavelengths, with some equatorial exceptions chosen so as to be available to telescopes from both hemispheres. This is likely to change to some extent, in parallel with the need to open up the southern FIR/Submm sky in anticipation and support of ALMA, and the Akari Deep Field South (ADFS, near the South Ecliptic Pole) multi-wavelength surveys are a first example of this trend. Originally chosen because, due to its orbit, the Akari satellite was going to spend quite some time around the polar sky, the field was later imaged at most "conventional" wavelengths and exploratory efforts have now being undertaken in the FIR/Sub-mm both with BLAST and with LABOCA@APEX \& AzTEC@ASTE, waiting for Herschel. A similar process could certainly be set in motion if and when a polar FIR/Submm observing facility is put in place.
\end{itemize}
\section{Conclusions}
The opportunity to exploit the high FIR/Submm atmospheric transparency at Dome C to carry out high angular resolution extragalactic surveys in this elusive wavelength range certainly ranks very high within Dome C astronomical science cases.
Installing state-of-the-art 200/350/450 $\mu$m filled bolometer arrays at the focal plane of an ALMA-like 12\,m winterized antenna would provide order-of-magnitude improvements in angular resolution with respect to Herschel at the same wavelengths, allowing to probe more deeply the sources making up the bulk of the CIRB up to the highest redshifts.

It appears, however, that while 350 \& 450 $\mu$m surveys could be efficiently carried out during most of the polar winter, a 12\,m antenna might not be enough in order to suitably exploit the more limited openings of the 200 $\mu$m atmospheric window, but that one would like to have access to a larger, notionally 25\,m, telescope.

More realistically, a 12\,m antenna following the ASO concept to be deployed at Dome C within a relatively short time frame would arguably be the best pathfinder for any larger-size astronomical facility on the Antarctic plateau, establishing it as a first-rate observing site and enabling ground-breaking science in is own right.

\end{document}